\documentclass[
aps,
pra,
superscriptaddress,
noshowpacs,
onecolumn,
preprint,
12pt,
floatfix
]
{revtex4-2}
\usepackage{graphicx}
\usepackage[colorlinks,
            linkcolor=blue,
            citecolor=blue,
            anchorcolor=blue,
            urlcolor=blue,
            ]{hyperref}
\usepackage{booktabs}
\usepackage{leftidx}
\usepackage{subfigure}
\usepackage{amsmath}
\usepackage{amssymb}
\usepackage{epstopdf}
\usepackage{amsmath}
\usepackage{lineno}
\usepackage{bm}
\usepackage{siunitx}
\DeclareSIUnit\dBm{dBm}

\newcommand{\figref}[1]{\mbox{Fig.~\ref{#1}}}
\newcommand{\figpanel}[2]{Fig.~\hyperref[#1]{\ref*{#1}(#2)}}
\newcommand{\figpanels}[3]{Fig.~\hyperref[#1]{\ref*{#1}(#2)-(#3)}}
\newcommand{\figpanelNoPrefix}[2]{\hyperref[#1]{\ref*{#1}(#2)}}

\newcommand{\ie}{{\em i.e.}}

\begin{document}
\title{Engineering the Environment of a Superconducting Qubit with an Artificial Giant Atom}
\author{Jingjing Hu}
\thanks{These two authors contributed equally to this work.}
\affiliation{Tencent Quantum Laboratory, Tencent, Shenzhen, Guangdong 518057, China}
\author{Dengfeng Li}
\thanks{These two authors contributed equally to this work.}
\affiliation{Tencent Quantum Laboratory, Tencent, Shenzhen, Guangdong 518057, China}
\author{Yufan Qie}
\affiliation{Tencent Quantum Laboratory, Tencent, Shenzhen, Guangdong 518057, China}
\author{Zelong Yin}
\affiliation{Department of Applied Physics, Stanford University, Stanford, California 94305, USA}
\author{Anton Frisk Kockum}
\affiliation{Department of Microtechnology and Nanoscience (MC2), Chalmers University of Technology, 412 96 Gothenburg, Sweden}
\author{Franco Nori}
\affiliation{RIKEN Center for Quantum Computing, Wakoshi, Saitama 351-0198, Japan}
\affiliation{Theoretical Quantum Physics Laboratory, RIKEN Cluster for Pioneering Research, Wako-shi, Saitama 351-0198, Japan}
\affiliation{Physics Department, The University of Michigan, Ann Arbor, Michigan 48109-1040, USA}
\author{Shuoming An}
\email{shuomingan@tencent.com}
\affiliation{Tencent Quantum Laboratory, Tencent, Shenzhen, Guangdong 518057, China}

\date{\today}

\begin{abstract}

In quantum computing, precise control of system-environment coupling is essential for high-fidelity gates, measurements, and networking. 
We present an architecture that employs an artificial giant atom from waveguide quantum electrodynamics to tailor the interaction between a superconducting qubit and its environment. 
This frequency-tunable giant atom exhibits both frequency and power selectivity for photons: when resonant with the qubit, it reflects single photons emitted from the qubit while remaining transparent to strong microwave signals for readout and control. 
This approach surpasses the Purcell limit and significantly extends the qubit's lifetime by ten times while maintaining the readout speed, thereby improving both gate operations and readout.
Our architecture holds promise for bridging circuit and waveguide quantum electrodynamics systems in quantum technology applications.

\end{abstract}

\maketitle


\section*{Main Text}
\subsection*{Introduction}

Superconducting qubits offer exceptional design flexibility, prolonged coherence time, and impressive scalability, making them competitive candidates for implementing scalable quantum computation~\cite{ladd2010quantum}. 
To execute complex quantum computational tasks that require numerous qubits, such as quantum error correction~\cite{lidar2013quantum}, it is crucial to perform fast readout and gate operations while maintaining long coherence times. 
However, fast readout and control demand large coupling between the qubits and their surrounding environment, limiting qubit lifetime.
For fast control, decreasing the coupling comes at the expense of having to increase driving power, which will hit a ceiling due to heating issues~\cite{krinner2019engineering}.
To decrease qubit decay from fast readout, a common strategy is to introduce an additional filter mode, improving the Purcell limit~\cite{reed2010fast,jeffrey2014fast}.
However, with recent advancements in coherence times for planar superconducting transmon qubits, now exceeding \SI{500}{us}~\cite{place2021new,wang2022towards,bal2024systematic}, even an improved Purcell limit may become the main coherence bottleneck if fast readout is required. 
This is particularly relevant for practical applications where increasing the readout coupling is preferred to eliminate the need for a quantum-limited amplifier~\cite{chen2023transmon}.
Such an approach is beneficial within large-scale chips~\cite{li2023optimizing}, as it helps to circumvent bandwidth and power-saturation constraints.

Here, to address these challenges, we redesign a Josephson quantum filter (JQF)~\cite{kono2020breaking,koshino2020protection} in a giant-atom~\cite{kockum2021quantum} form in a waveguide that is used for both readout and control. 
The giant-atom JQF, illustrated in \figref{fig:Picture}, is a transmon~\cite{koch2007charge}, whose frequency can be tuned in situ, strongly coupled to the readout/control waveguide at two points. 
The distance in the waveguide between the two coupling points is comparable to the photonic wavelength in the waveguide, providing opportunities to control the qubit's coupling to the environment through interference effects~\cite{kockum2021quantum,wang2024realizing}.
The giant-atom JQF exhibits both frequency and power selectivity; when tuned into resonance with the qubit, the JQF will only reflect the ``single" photon emitted from the qubit in a decay process (effectively increasing the coherence time) without affecting the strong microwave tone (consisting of many photons) used for readout/control~\cite{mirhosseini2019cavity,kono2020breaking}. 

Our giant-atom JQF has several advantages over previous designs~\cite{kono2020breaking}, where a JQF was placed solely on the control line, which does not eliminate leakage through the readout waveguide. 
While one can trade lower coupling of the qubit to the control line against higher driving power up to a heating ceiling~\cite{krinner2019engineering}, the Purcell limit for the readout is not as directly addressable: increasing readout microwave power results in the excitation of the qubit outside the computation space or the transmon ionization~\cite{sank2016measurement, yin2022shortcuts, dumas2024unified}. 
Therefore, a sufficiently large readout resonator-waveguide coupling $\kappa_{\rm r}$ is necessary for fast readout, particularly when considering scenarios without a quantum-limited amplifier. 
Yet, with the common Purcell filter, as demonstrated in Ref.~\cite{chen2023transmon}, qubit lifetime $T_1$ could be limited to below \SI{7}{\micro\second} for $\kappa_{\rm r}/2\pi>\SI{11}{\mega\hertz}$. 
With the giant-atom JQF, the qubit's excited state approximates a non-decaying dark state that is shielded from decay through the readout/control waveguide, without being significantly limited by $\kappa_{\rm r}$. 
Indeed, the multiple coupling points of the giant-atom JQF increases the effective JQF-waveguide coupling, which improves the dark-state approximation.

Another advantage of the giant-atom JQF is its scalability to a multi-qubit setup. 
Its high resilience against waveguide mismatches could mean that arranging more qubits and giant-atom JQFs in a sequence along an waveguide for multiplexed readout is possible, while we still need individual controls.
Furthermore, the giant-atom configuration with a single waveguide for both readout and control simplifies the system and enhances its performance by only using one JQF to control the total environment which the qubit is coupled to, and eliminates the possibility of frequency-mismatch errors for multiple JQFs along one waveguide.

\subsection*{Setup and Working Principles}

\begin{figure}[!htb] 
\centering
\includegraphics[width=1\textwidth]{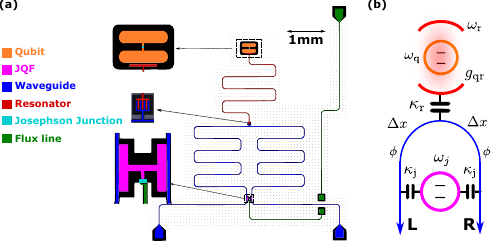} 
\caption{
\textbf{Optical micrographs and schematics of the giant-atom Josephson quantum filter.}  
(a) False-color optical micrograph of the giant-atom JQF chip. 
The qubit is a fixed-frequency transmon (orange) coupled to a half-wavelength coplanar waveguide readout resonator (red), connected to a meandered waveguide (blue) through a digital finger capacitor. 
The waveguide connects to the JQF (pink), a frequency-tunable transmon, at two separate coupling points. 
A wire-bonding bridge, not shown here, connects the flux line (green) for the JQF, and flux-trapping holes prevent coherence disturbance from environmental magnetic bias. 
(b) Schematic illustration of the JQF, featuring qubit frequency $\omega_{\rm q}$, resonator frequency $\omega_{\rm r}$, JQF frequency $\omega_{\rm j}$, readout resonator damping rate $\kappa_{\rm r}$, JQF decay rate $\kappa_{\rm j}$ at each coupling point, qubit-resonator coupling $g_{\rm qr}$, and the distance $\Delta x$ between each coupling point and the qubit-readout resonator coupling point.
The condition $\phi = \omega_{\rm q} \Delta x / v_{\rm p} = \pi$ is satisfied, where $v_{\rm p}$ represents the phase velocity of photons with frequency $\omega_{\rm q}$ in the waveguide.
} 
\label{fig:Picture} 
\end{figure}

Our device consists of a fixed-frequency transmon qubit connected to a readout resonator for dispersive readout; the resonator is coupled to the midpoint of the waveguide of a giant-atom JQF, as illustrated in \figpanel{fig:Picture}{a}. 
The giant-atom JQF is symmetrically coupled to the waveguide at two points, with both coupling points located at a distance $\Delta x$ from the midpoint. 
This distance $\Delta x$ satisfies the condition: $\phi = \omega_{\rm q} \Delta x / v_{\rm p} = \pi$, where $\omega_{\rm q}$ is the qubit frequency and $v_{\rm p}$ is the phase velocity of photons with frequency $\omega_{\rm q}$ in the waveguide.

Here, we present an intuitive understanding of the working principles. 
We consider a simplified model, where the qubit is described as being effectively coupled to the waveguide with a strength $\kappa_{\rm q}$. 
This effective coupling results from the resonator coupling to both the qubit and the waveguide, and can be estimated from the Purcell limit to be about $2\pi \times \SI{15}{\kilo\hertz}$. 
In the Supplementary Material, we analyze the full dynamics of our setup, using the SLH formalism~\cite{combes2017slh, kockum2018decoherence} for cascaded quantum systems.
In the giant-atom JQF configuration depicted in \figpanel{fig:Picture}{b}, the JQF is coupled to the waveguide at two points, with decay rate of $\kappa_{\rm j}$ through each point. 
When the JQF frequency matches the qubit frequency $\omega_{\rm q}$ and the distance $\Delta x$ between the qubit and the JQF satisfies the condition $\phi = \Delta x\omega_{\rm q}/v_p \approx \pi$, the system exhibits specific properties. 

Firstly, as shown in Ref.~\cite{kannan2020waveguide}, the waveguide mediates a distance-dependent exchange interaction between the qubit and the giant-atom JQF. 
The strength of this interaction is $J_{i,j} = 2 \pi g_{i} g_{j} \omega_{\rm q} \sin(\phi)$, where $g_{i}$ and $g_{j}$ denote the coupling strengths of the single emitters (the qubit and the JQF), and the mode within the waveguide at the qubit frequency.
Due to the symmetric position of the qubit relative to the two coupling points of the giant atom, this exchange interaction is zero for integer multiples of $\phi = \pi$. 
The final Hamiltonian therefore only consists of the two eigensystems of the qubit and the giant atom JQF.

Secondly, for the correlated dissipation introduced by the waveguide, the symmetry means that we only need to analyze one port. 
Arriving at the $\bf{R}$ port, a photon emitted from the JQF at the $\bf{L}$ port has accumulated a phase of $2\phi = 2\pi$, with an amplitude of $\sqrt{\frac{\kappa_{\rm j}}{2}}\sigma_{\rm j}^{-}$. 
Similarly, a photon emitted from the qubit accumulates a phase of $\phi = \pi$ and has an amplitude of $-\sqrt{\frac{\kappa_{\rm q}}{2}}\sigma_{\rm q}^{-}$. 
The amplitude of the photon emitted from the JQF at the $\bf{R}$ port is $\sqrt{\frac{\kappa_{\rm j}}{2}}\sigma_{\rm j}^{-}$. 
These three decay processes interfere at the $\bf{R}$ port and will be dissipated there if the waveguide impedance is matched. 
The corresponding dissipator is $L_{R} = \sqrt{2\kappa_{\rm j}}\sigma_{\rm j}^{-}-\sqrt{\frac{\kappa_{\rm q}}{2}}\sigma_{\rm q}^{-}$. 
A non-trivial dark state $|D\rangle$ of this dissipator ($L_{R}|D\rangle = 0$) in the $|\rm{qubit},\rm{JQF}\rangle$ basis is straightforward to identify: $|D\rangle \propto \sqrt{\frac{\kappa_{\rm q}}{2}}|01\rangle+\sqrt{2 \kappa_{\rm j}}|10\rangle$. 
In our design, we have a super-radiant decay rate of $4\kappa_{\rm j}/2\pi \approx$\SI{50}{\mega\hertz} for the JQF. 
Given that $4\kappa_{\rm j}/\kappa_{\rm q}\approx 3000$, we obtain $\left|D\rangle\approx|10\right\rangle$, corresponding to the qubit's excited state.

In an optimal parameter configuration, the qubit can maintain its approximately dark excited state even when strongly coupled to the environment, thereby enabling fast readout.
Without a functioning giant atom JQF, the parameters of our device would lead to a Purcell-limited qubit lifetime of \SI{9}{\micro\second}.
The readout signal, which has a frequency significantly different from the qubit and giant atom frequencies, is not significantly affected by the presence of the giant-atom JQF.
Furthermore, the high-power control signal employed for single-qubit gates, which is estimated to be over 100 times greater than the weak power used for giant-atom spectroscopy, saturates the giant atom, effectively negating any influence on qubit control.
Consequently, this configuration allows for high-speed readout and control of a qubit while simultaneously preserving its extended lifetime.



\subsection*{Spectrum and saturation of the giant-atom JQF}
\begin{figure}[!htb] 
\centering
\includegraphics[width=1\textwidth]{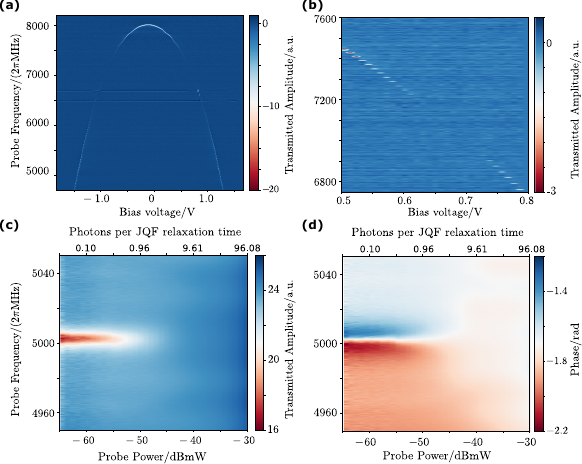} 
\caption{
\textbf{Basic properties of the giant-atom JQF.}  
(a) Giant-atom JQF frequency spectrum.
We adjust the giant-atom JQF's eigenfrequency by tuning the magnetic flux through the Josephson junction loop via a bias voltage (x-axis).
The amplitude variation in the microwave transmission reveals the giant atom's eigenfrequency under weak probe power.
We subtracted signals without the giant-atom spectrum as background for this figure.
Horizontal lines are from readout resonators on the chip.
(b) Expanded view of the sub-radiant spectrum range.
The sub-radiant configuration occurs when the frequency $\omega_{\rm j}$ and the distance $2\Delta x$ between the two giant-atom coupling points meet $2\phi = 2\Delta x \omega_{\rm j}/v_{\rm p}= 3\pi$.
Microwave response disappears here due to destructive interference from the two coupling points.
(c, d) Power-saturation characteristics of the giant atom.
The lower x-axis shows the probe power of the vector network analyzer, with a line microwave attenuation of about \SI{-80}{\dB}.
The upper x-axis indicates the corresponding photon number passing through the JQF during one relaxation time.
Both amplitude and phase spectral features disappearing when the probe power surpasses \SI{-40}{\dBm}, indicating the two-level nature of the system.
} 
\label{fig:Spectrum} 
\end{figure}

We now turn to characterization of our giant-atom JQF device depicted in \figref{fig:Picture}.
First, we explore its frequency spectrum to understand its properties and determine the working flux bias; see \figpanel{fig:Spectrum}{a,b}. 
The eigenfrequency of the giant atom can be tuned by applying a bias voltage, x axis in \figpanel{fig:Spectrum}{a,b}, through the flux line. 
In this experiment, we probe the system by applying a microwave tone of \SI{-55}{\dBm} with the probe frequency displayed on the y-axis in all panels of \figref{fig:Spectrum}, originating from the vector network analyzer (please refer to the Supplementary Material for a comprehensive description of the measurement setup). 
This corresponds to \SI{-135}{\dBm} arriving at the chip or a photon flux of $\dot{n}=\SI{e6}{\per\second}$. 
From \figpanel{fig:Spectrum}{c}, we can estimate the damping rate $\kappa$ of the tested JQF to be approximately $2\pi\times$\SI{5}{\mega\hertz}. 
This implies that $\dot{n}/\kappa=0.3$ photons per relaxation time will reach the JQF.

When the probe frequency matches the eigenfrequency $\omega_{\rm j}$ of the giant atom, we observe an abrupt amplitude and phase change in the transmitted probe amplitude in \figpanel{fig:Spectrum}{a}.
However, when the frequency $\omega_{\rm j}$ and the distance $2\Delta x$ satisfy $2\phi = 2\Delta x \omega_{\rm j}/v_{\rm p} = (n+1/2) 2\pi$ with $n$ integer, a sub-radiant state emerges, where no change is observed in the transmission of resonant microwaves, as shown in \figpanel{fig:Spectrum}{b}. 
Due to the band-pass filter on the microwave cables, we observe this around \SI{7050}{\mega\hertz}. 
The sub-radiance of this state is due to destructive interference of the emission from the two coupling points of the giant atom, similar to an artificial atom in front of a mirror interfering with its own reflected emission~\cite{hoi2015probing}. 
In contrast, when the distance aligns with the photonic wavelength at the eigen-frequency of the giant atom or $\phi = n \times 2\pi$, a super-radiant state of the giant atom is induced, effectively resulting in an enhanced coupling of the giant atom to the environment.

In \figpanel{fig:Spectrum}{c,d}, we characterize the two-level nature of the giant atom by measuring its power saturation.
When the applied microwave power surpasses \SI{-40}{\dBm} or 9.6 photons per JQF relaxation time, we see that the giant atom spectrum vanishes from the transmitted microwave signal, since it only can influence on the order of one photon per relaxation time.
This implies that the presence of the giant atom will not have a significant impact on the system when subjected to intense control or readout driving, which has a maximum power of approximately \SI{-10}{\dBm} in our experimental setup.


\subsection*{Improving qubit relaxation time}

\begin{figure}[!htb] 
\centering
\includegraphics[width=1\textwidth]{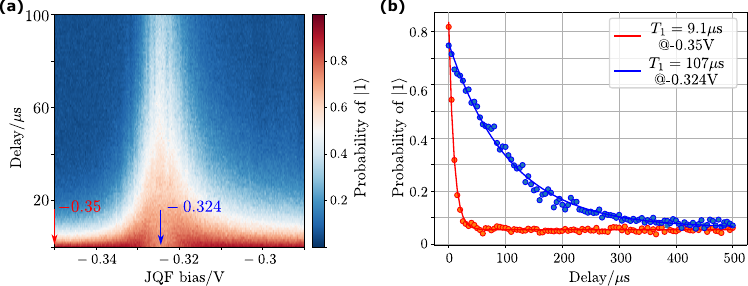} 
\caption{
\textbf{Qubit relaxation for different giant-atom bias voltages.}
(a) Qubit relaxation lifetime $T_1$ for various giant-atom JQF bias voltages.
In each measurement, we set the giant-atom JQF bias, excite the qubit to the $|1\rangle$ state, and measure the probability of the qubit remaining in the excited state after a varying time delay.
After a \SI{100}{\micro\second} waiting period for qubit reset, the bias voltage changes, and the delay scan restarts.
The qubit lifetime extends when the JQF bias is tuned on resonance with the qubit, at about \SI{-0.324}{\volt} in our device.
Around this point, the initial $|1\rangle$ probability is smaller due to insufficient reset time, but it does not affect the $T_1$ results.
(b) Comparison of the best and worst $T_1$ times for different JQF bias voltages.
The $|1\rangle$ state probability with varying delays is fitted using an exponential decay model.
The JQF can prolong the qubit lifetime $T_1$ from \SI{9.1}{\micro\second} at the JQF idle point (far detuned from the qubit) to \SI{107}{\micro\second} at the JQF working point, a more than tenfold increase.
Each data point is the average of 1000 repetitions.
} 
\label{fig:Relaxation} 
\end{figure}

To evaluate the giant-atom JQF's effectiveness in preventing qubit decay, we measure the qubit lifetimes $T_1$ at various giant-atom frequencies.
The qubit frequency in this tested chip is $\omega_{\rm q}/2\pi = \SI{5.011}{\giga\hertz}$. 
As shown in \figref{fig:Relaxation}, when the giant atom is tuned from being off resonance to being in resonance with the qubit frequency using an appropriate bias voltage, the qubit's lifetime increases from \SI{9.1}{\micro\second} to \SI{107}{\micro\second}, a more than tenfold improvement. 
We also estimate the qubit lifetime without a functional giant-atom JQF, based on the Purcell limit of the readout resonator, to be approximately \SI{10}{\micro\second}, consistent with our experimental observations. 

The chip tested in this study was fabricated using a sapphire substrate and a body-centered cubic tantalum film~\cite{potts2001cmos,place2021new,Li2024MinimizingKI}. 
In the qubit with the giant-atom JQF, the observed $T_1$ is more than \SI{100}{\micro\second}, which is still lower than the results reported in Ref.~\cite{place2021new}. 
This difference may be due to two factors. 
First, the coupling between the JQF and the waveguide spans \SI{200}{\micro\meter} in the design, introducing an uncertainty in $\phi$. 
Second, the approximated qubit excited state also contains a small portion of the giant-atom excited state, which is smaller in size and has a larger energy participation at the lossy metal-air interface.
Despite these imperfections, the JQF still enhances the qubit lifetime tenfold. 
This implies that the giant-atom JQF design possesses a relatively large tolerance, potentially rendering it suitable for multi-qubit architectures.
Moreover, the asymmetry profile of $T_1$ we observe in \figpanel{fig:Relaxation}{a} is a result of the mismatch of $\phi$ to the ideal $\pi$. 
However, this mismatch will not significantly degrade the protection effect within a broad range. 
For more numerical details~\cite{johansson2012qutip} on the robustness to this mismatch, see the Supplementary Material.


\subsection*{Performance enhancements for single-qubit operations}

\begin{figure}[!htb] 
\centering
\includegraphics[width=0.9\textwidth]{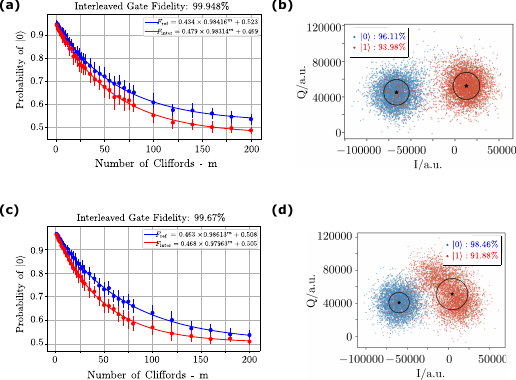} 
\caption{
\textbf{Comparison of single-qubit operations with and without a working giant-atom JQF.}
(a) Interleaved randomized benchmarking (IRB) of a single-qubit $X$ gate and (b) data from single-qubit readout; in both panels, the giant-atom bias is optimized for maximum $T_1 = \SI{75}{\micro\second}$. 
(c, d) Same as (a,b), but with the giant atom detuned such that $T_1 = \SI{8}{\micro\second}$. 
From (c) to (a), the single-qubit gate error decreases by a factor of six, from $0.33\%$ to $0.052\%$. 
In the RB sequence, all $\pm X/2(Y/2)$ and $\pm X(Y)$ gates have a duration of \SI{50}{\nano\second}; the $\pm X/2(Y/2)$ gate uses half of the amplitude of the $\pm X(Y)$ gate and remains unoptimized.
Each data point is obtained by averaging the results from 30 random circuits for each number $m$ of Clifford gates; each circuit is executed 1000 times.
In panel (b), with the qubit lifetime extended by the active giant-atom JQF, the blobs from the dispersive readout (5000 data points are collected for each qubit state) appear more concentrated than in panel (d), suggesting less qubit decay during the \SI{900}{\nano\second} rectangular readout pulse. 
The readout power is maintained at \SI{-18}{\dBm} with room temperature electronics, and all other parameters are kept constant throughout (b,d), except for the giant-atom bias.
} 
\label{fig:T1_SQ} 
\end{figure}

Merely demonstrating that the qubit's lifetime can be extended is insufficient to assert that the giant-atom JQF can be utilized in a practical quantum computing device. 
We therefore now turn to characterizing the performance improvement due to the giant-atom JQF for two single-qubit operations: an $X$ gate and qubit readout.
The results of the characterization are presented in \figref{fig:T1_SQ}.
For the single-qubit $X$ gate, we employ interleaved randomized benchmarking (IRB)~\cite{magesan2012efficient} to estimate the gate fidelity.
The local gate set includes $\{I, \pm X/2, \pm Y/2, \pm X, \pm Y\}$, with $X$ as the interleaved gate. 
For different giant-atom biases, we conduct the IRB and find that the error per $X$ gate can be reduced from $0.33\%$ at the giant-atom JQF idle point [\figpanel{fig:T1_SQ}{c}] to $0.052\%$ at the giant-atom JQF working point [\figpanel{fig:T1_SQ}{a}]. 
This suggests that the single-qubit gate error decreases by a factor of six. 
We note that the improvement ratio of the single-qubit gate error is not as significant as that of $T_1$, although the coherence limit for gate fidelity would imply that they should be the same~\cite{abad2022universal}. 
We attribute this discrepancy to three reasons: firstly, the decoherence $T_2 = (\frac{1}{2T_1}+\frac{1}{T_\phi})^{-1}$ will not increase as $T_1$, as the dephasing $T_{\phi}$ will not be improved by the JQF; secondly, the remaining control error and leakage to higher qubit levels do not decrease; thirdly, as shown in Ref.~\cite{kono2020breaking}, the JQF also slightly distorts the control waveform, resulting in a larger control error than the coherence limit.

In \figpanel{fig:T1_SQ}{b,d}, we compare the performance differences in dispersive readout. 
With the readout power and all other readout parameters kept constant, the qubit is not excited by the strong readout power~\cite{sank2016measurement,yin2022shortcuts,khezri2023measurement}.
We observe that the readout blobs become more concentrated when the giant-atom JQF is tuned to the working point in \figpanel{fig:T1_SQ}{b}, while there is a ``tail'' of $|1\rangle$ when the JQF is at the idle point in \figpanel{fig:T1_SQ}{d}.
We attribute this increased concentration in \figpanel{fig:T1_SQ}{b} to the reduced qubit decay during the \SI{900}{\nano\second} rectangular readout pulse.
However, we do not observe a significant difference in readout fidelity between the two cases. 
Indeed, the $|1\rangle$ readout fidelity increases by about 2 percent when using the giant-atom JQF, but the $|0\rangle$ readout fidelity instead degrades by about 2 percent. 
This result could be due to two factors. 
First, the increased reset time required for optimal readout when the qubit has a longer $T_1$. 
In both experiments, we keep the reset time constant at \SI{500}{\micro\second}. 
Second, when the JQF is at the working point, the non-trivial dark state of the qubit-JQF combined system contains a small amount of the qubit ground state.
These factors mean the state-preparation error is larger with a working JQF, a phenomenon also observed in previous study~\cite{kono2020breaking}.


\subsection*{Discussion}

In conclusion, we have explored the interaction between WQED and CQED systems, demonstrating successful control of the qubit's environment using a giant-atom JQF. 
This has resulted in a more than tenfold increase in qubit lifetime, exceeding \SI{100}{\micro\second}. 
Influenced by the giant-atom JQF, the single-qubit gate error has been reduced by six times, and the dispersive readout performance has also been altered. 
This ability to manipulate the environment of a superconducting qubit with high tunability~\cite{kallush2022controlling, wang2024realizing} opens up new possibilities for applications in quantum information processing and quantum networking.

A direct application of the giant-atom JQF lies in its potential for fast qubit reset. 
Traditional passive reset methods usually require 5--10 times the qubit lifetime to complete~\cite{mcewen2021removing, zhou2021rapid}. 
By waiting with a detuned JQF and employing circuits with a functional JQF, we can save at least approximately \SI{400}{\micro\second} for each experimental run.

In this study, we implemented a giant-atom JQF in a transmissive setup and explored its application in a reflective setup (see Supplementary Material). 
The reflective configuration directs all photon emissions to a specific location, allowing the JQF to act as a tunable coupler between the qubit and the waveguide environment.
As proposed in Ref.~\cite{cirac1997quantum}, this facilitates high-fidelity quantum state transfer and entanglement distribution among distant chips. 
Notably, unlike recent experiments~\cite{zhong2021deterministic} utilizing inductive couplers, the JQF interacts solely with the single photon emitted from the qubit, without being part of the qubit mode. 
This feature allows the qubit frequency to remain constant while adjusting the coupler.


\bibliography{MRF}


\section*{Acknowledgments}

We would like to express our sincere gratitude to X. Wang for his comments on the manuscript, and to S.Y. Zhang for his support throughout this project.


\subsection*{Author contributions}

S.M.A., Y.F.Q., and Z.L.Y.: Conceptualization and methodology.
J.J.H., D.F.L., and S.M.A.: Investigation.
S.M.A., A.F.K., and F.N.: Writing.


\subsection*{Competing Interests Statement}

The authors declare no competing interests.


\subsection*{Data And Code Availability}

All data needed to evaluate the conclusions in the paper are present in the paper or the supplementary materials.


\newpage

\section*{Supplementary Materials}

\renewcommand{\thefigure}{S\arabic{figure}}
\setcounter{figure}{0}
\renewcommand{\thetable}{S\arabic{table}}
\setcounter{table}{0}



\subsection*{SLH calculation}
\label{sec:SLH}

In this section, we provide comprehensive derivations for all master equations related to small and giant atoms as discussed in the main text. 
We strive to make these derivations self-contained. 
For further details, please refer to the reviews of the method employed here, the SLH formalism for cascaded quantum systems, in Ref.~\cite{combes2017slh, kockum2018decoherence}.

An open quantum system with $n$ input-output ports can be described by an SLH triplet $G = (\mathbf{S}, \mathbf{L}, H)$, where $\mathbf{S}$ is an $n\times n$ scattering matrix, $\mathbf{L}$ is an n$\times$1 vector describing the system's coupling to the environment at the input-output ports, and $H$ denotes the system's Hamiltonian. 
Once the SLH triplet $G=(\mathbf{S}, \mathbf{L}, H)$ for a system is obtained, the master equation for that system is given by
\begin{equation}
\dot{\rho} = -i [H, \rho] + \sum_{j=1}^{n} \mathcal{D}\left[L_{j}\right] \rho,
\end{equation}
where $\rho$ is the system density matrix and $\mathcal{D}[X] \rho=X \rho X^{\dagger}-\frac{1}{2} X^{\dagger} X \rho-\frac{1}{2} \rho X^{\dagger} X$ denotes Lindblad operators. The output from port $j$ of the system is simply represented by $L_{j}$.
It is important to note that the SLH formalism is based on the same physical assumptions as the standard Lindblad master equation, \ie, weak coupling and the Markov approximation. 
Furthermore, the SLH formalism, as presented thus far, also necessitates that the fields connecting various systems propagate in linear, dispersionless media, and that the propagation time is negligible.
These conditions are satisfied in the setups we consider.

To consolidate SLH triplets of cascaded quantum systems into a single triplet characterizing the entire setup, we employ two composition rules: the series product $\triangleleft$ and the concatenation product $\boxplus$. 
If two systems, represented by $G_1$ and $G_2$, are combined in series such that output from $G_1$ is used as input to $G_2$, the resulting total SLH triplet is determined by the series product:
\begin{equation}
G_2 \triangleleft G_1=\left(\mathbf{S}_2 \mathbf{S}_1, \mathbf{S}_2 \mathbf{L}_1+\mathbf{L}_2, H_1+H_2+\frac{1}{2 i}\left[\mathbf{L}_2^{\dagger} \mathbf{S}_2 \mathbf{L}_1-\mathbf{L}_1^{\dagger} \mathbf{S}_2^{\dagger} \mathbf{L}_2\right]\right).
\end{equation}
If these two systems are combined in parallel, the resulting total SLH triplet is determined by the concatenation product:
\begin{equation}
G_1 \boxplus G_2=\left(\left[\begin{array}{cc}
\mathbf{S}_1 & 0 \\
0 & \mathbf{S}_2
\end{array}\right],\left[\begin{array}{l}
\mathbf{L}_1 \\
\mathbf{L}_2
\end{array}\right], H_1+H_2\right).
\end{equation}

When defining the $\mathbf{S}$ matrix, it is crucial to consider the frequency-dependent nature of the phase shift. 
For the $\mathbf{L}$ matrix, in the case of a qubit coupled to a symmetric waveguide, the emission in one direction should be given by $\sqrt{\frac{\kappa_{\rm q}}{2}}\sigma^{-}$, where $\kappa_{\rm q}$ is the total decay rate of the qubit into the waveguide (via the readout resonator, in our case). 
Since there are two emission directions, $\kappa_{\rm q}$ is divided by 2.
Lastly, when formulating the $H$ matrix, it is essential to ensure that the Hamiltonian for the system is expressed only once.

\begin{figure}[!htb] 
\centering
\includegraphics[width=0.9\textwidth]{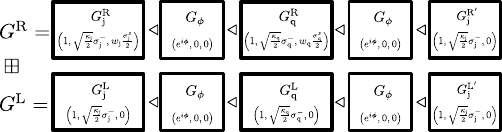} 
\caption{
\textbf{SLH representation of the giant-atom JQF.}
The open system has two ports, $G^{\rm L}$ and $G^{\rm R}$, which are combined in parallel.
The $G^{\rm L(R)}$ each contains several SLH triplet in series. 
}
\label{fig:SLH} 
\end{figure}
In the giant-atom JQF case, the system is depicted in Fig.~1 in the main text.
The total SLH triplet is given by
\begin{equation}
G_{\rm tot}=G_{\rm R}\boxplus G_{\rm L},
\end{equation}
where all symbols are defined in \figref{fig:SLH}.
When $\omega_{\rm j} = \omega_{\rm q} = \omega_{0}$ and $\phi = \Delta x\cdot\omega_0/v_{p} =\pi$, there is no exchange interaction between the qubit and the giant-atom JQF, resulting in the following SLH triplet:
\begin{equation}
\left\{\begin{array}{l}
\mathbf{S}=\left(\begin{array}{ll}
1 & 0 \\
0 & 1
\end{array}\right) \\
\mathbf{L}=\left(\sqrt{2 \kappa_{\rm j}} \sigma_{\rm j}^{-}-\sqrt{\frac{\kappa_{\rm q}}{2}} \sigma_{\rm q}^{-}, \sqrt{2 \kappa_{\rm j}} \sigma_{\rm j}^{-}-\sqrt{\frac{\kappa_{\rm q}}{2}} \sigma_{\rm q}^{-}\right) \\
H=\frac{1}{2} w_{\rm j} \sigma_{\rm j}^z+\frac{1}{2} w_{\rm q} \sigma_{\rm q}^z.
\end{array}\right.
\end{equation}
We observe that the dissipators for the right and left propagation directions are identical. 
The system possesses one non-trivial dark state, which satisfies $\mathbf{L}|D\rangle=\mathbf{0}$:
\begin{equation}
|D\rangle \propto\sqrt{\frac{\kappa_{\rm q}}{2}}|01\rangle+\sqrt{2 \kappa_{\rm j}}|10\rangle.
\end{equation}
When $\kappa_{\rm j}\gg \kappa_{\rm q}$, the dark state for both situations corresponds to the first excited state of the qubit: $|10\rangle$.


\subsection*{Purcell limit}

We first consider a qubit-resonator system without the inclusion of a Purcell filter. 
Within the one-excitation subspace of the Hilbert space, the Hamiltonian of this combined system can be written as
\begin{linenomath*}
\begin{equation}
    H_{\rm qr}=\begin{pmatrix}
      \omega_{\rm q} & g_{\rm qr} \\
      g_{\rm qr}^{*} & \omega_{\rm r}-i\kappa_{\rm r}/2
    \end{pmatrix}.
    \label{eq:H_qr}
\end{equation}
\end{linenomath*}
In the giant-atom JQF configuration that we examine in this work, we have the qubit frequency $\omega_{\rm q}/2\pi=$\SI{5.011}{\giga\hertz}, the resonator frequency $\omega_{\rm r}/2\pi=$\SI{6.621}{\giga\hertz}, the qubit-resonator coupling $g_{\rm qr}/2\pi=$\SI{144}{\mega\hertz}, and the damping rate of the resonator to the waveguide environment is $\kappa_{\rm r}/2\pi=$\SI{2.1}{\mega\hertz}. 
Once all these parameters have been determined, we can diagonalize Eqn.~\ref{eq:H_qr} and obtain the complex qubit-mode eigenfrequency $\widetilde{\omega}_{\rm q}$. 
The Purcell limit of the qubit decay can then be calculated as $T_1 = 1/(-2\rm{Im}(\widetilde{\omega}_{\rm q}))=$\SI{9.7}{\micro\second}. 
It is clear that, in the absence of a giant-atom JQF, the qubit coherence would be significantly impacted by leakage into the waveguide through the resonator.


\subsection*{Chip fabrication}

Chips are fabricated on a \SI{330}{\micro\meter}-thick sapphire substrate. 
The substrate is cleaned in a piranha solution, heated in a load-lock chamber, and transferred to an UHV sputtering chamber. 
During deposition, the temperature is maintained at room temperature.
A \SI{10}{\nano\meter}-thick Nb buffer layer is deposited using a DC magnetron system, followed by a \SI{200}{\nano\meter}-thick bcc-Ta film~\cite{place2021new}. 
Readout lines and resonator circuits are patterned using direct laser writing and inductively coupled plasma etching techniques.
The chip is spin-coated with two layers of photoresist and exposed to \SI{100}{\kilo\electronvolt} electrons.
Samples undergo ultrasonic development to create Al/AlO$_x$/Al Josephson junctions using the bridge-free ``Manhattan style"~\cite{potts2001cmos}.
A plasma etch is employed to remove adhesive photoresist. 
Finally, lift-off is performed in a bath of remover, accompanied by sonication at \SI{80}{\celsius} for \SI{8}{hour} to complete the fabrication of the quantum chip.


\subsection*{Experimental setup}

\begin{figure}[!htb] 
\centering
\includegraphics[width=0.7\textwidth]{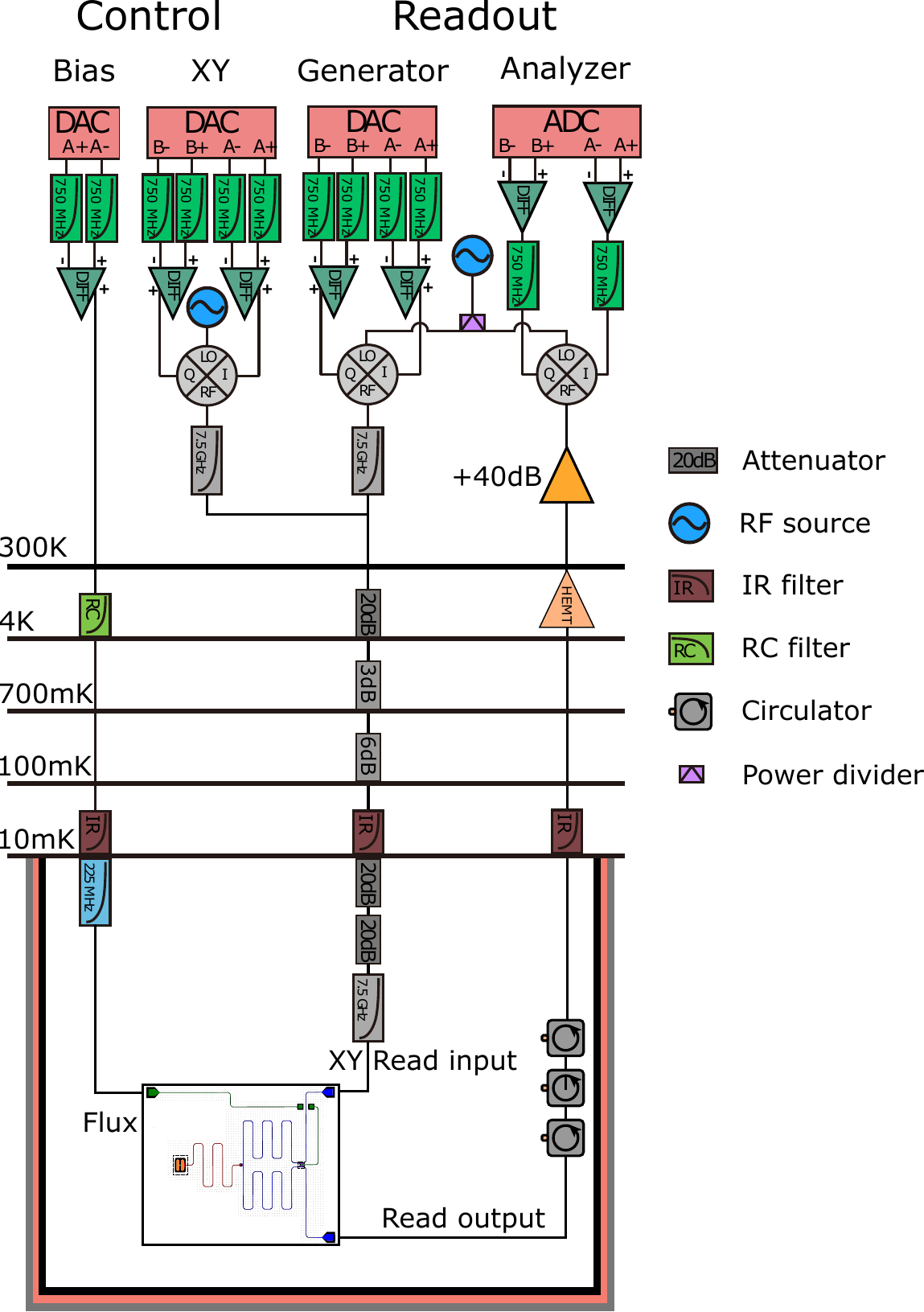} 
\caption{
\textbf{Complete wiring diagram for the experimental setup of the tested chip.}
The flux line is employed to control the flux bias of the giant-atom JQF. 
The XY control is executed through the readout input line. 
Abbreviations used in the diagram include digital-to-analog converter (DAC), analog-to-digital Converter (ADC), local oscillator (LO), radio frequency (RF), infrared (IR), and high electron mobility transistor (HEMT).
}
\label{fig:cryo} 
\end{figure}

We mount the JQF chip on the base plate of a dilution refrigerator and connect it to the control electronics located at room temperature, as illustrated in \figref{fig:cryo}.
We combine the XY control and readout input signals at room temperature.
Since the readout coupling is relatively large, the XY control speed is sufficient and a quantum-limited parametric amplifier is also not necessary.


\subsection*{Results for a reflective JQF chip}

\begin{figure}[!htb] 
\centering
\includegraphics[width=\textwidth]{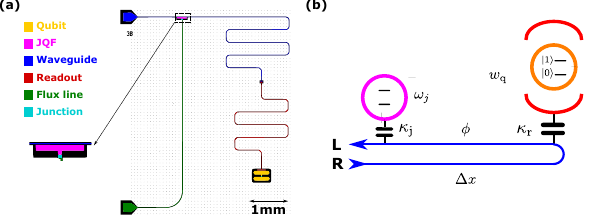} 
\caption{
\textbf{Optical micrographs and schematics of the reflective JQF chip.}  
(a) False-color optical micrograph of the reflective JQF chip. 
(b) Schematic of the reflective JQF setup.
} 
\label{fig:R_chip} 
\end{figure}

\begin{figure}[!htb] 
\centering
\includegraphics[width=0.7\textwidth]{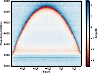} 
\caption{
\textbf{Frequency spectrum of the reflective JQF.}
} 
\label{fig:R_spectrum} 
\end{figure}

\begin{figure}[!htb] 
\centering
\includegraphics[width=\textwidth]{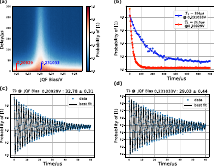} 
\caption{
\textbf{Low-temperature measurement results for the reflective JQF.}  
(a) Qubit relaxation lifetime $T_1$ for various JQF bias voltages. The figure is similar to Fig.~3 in the main text, but for the reflective JQF instead of the giant-atom JQF.
(b) The measured $T_1$ at the two JQF biases, as indicated in panel (a), reveals that the frequency corresponding to the lower bias voltage (\SI{0.20929}{\volt}) is approximately \SI{70}{\mega\hertz} higher than the one corresponding to the higher bias voltage (\SI{0.231033}{\volt}). 
The JQF frequency at the higher bias voltage closely aligns with the qubit frequency, both approximately at \SI{4431.9}{\mega\hertz}.
We measured 5 times longer $T_1$ when the JQF bias is set to the working point (red) than when it is off resonance (blue).
(c, d) The calculated dephasing times $T_\phi$ are obtained under the same two JQF bias voltages as mentioned above. 
In this case, $\frac{1}{T_2}=\frac{1}{2T_1}+\frac{1}{T_\phi}$, where $T_2$ represents the Ramsey decay time fitted from the experimental data, and $T_1$ is derived from previous measurements.
} 
\label{fig:Low temperature} 
\end{figure}

In the reflective JQF configuration (where the JQF is not a giant atom and the qubit is placed at the end of the waveguide), as illustrated in \figref{fig:R_chip}, we incorporate a JQF in the combined control and readout waveguide. Our design thus differs from that in Ref.~\cite{kono2020breaking}, where the JQF is only on the control waveguide.

When the frequency of the JQF is tuned to the qubit frequency $\omega_{\rm q}$, the amplitude of the photon emitted from the qubit, as observed from the position of the JQF, is $-\sqrt{\kappa_{\rm q}}\sigma_{\rm q}^{-}$.
Here, $\kappa_{\rm q}$ is the decay rate of the qubit, $\sigma_{\rm q}^{-}$ is the qubit's lowering operator, and the minus sign arises from the $\phi = \pi$ phase shift from the waveguide delay.
The amplitude of the photon emitted from the JQF is $\sqrt{\kappa_{\rm j}}\sigma_{\rm j}^{-}$, where $\kappa_{\rm j}$ is the decay rate of the JQF and $\sigma_{\rm j}^{-}$ is the lowering operator of the JQF.
These two photons interfere, resulting in an amplitude of $\mathbf{L} = -\sqrt{\kappa_{\rm q}}\sigma_{\rm q}^{-}+\sqrt{\kappa_{\rm j}}\sigma_{\rm j}^{-}$.
This photon is dissipated at the $50\ \Omega$ port, and $\mathbf{L}$ serves as the dissipator in the master equation.
In our design, the qubit is coupled to the waveguide through a readout resonator, and the decay rate $\kappa_{\rm q}/2\pi \approx 10\,\text{kHz}$ can be estimated from the Purcell limit.
Moreover, according to the design, the decay rate of the JQF is $\kappa_{\rm j}/2\pi  \approx 10\,\text{MHz}$.
Given that $\kappa_{\rm j}\gg\kappa_{\rm q}$, we obtain the dark state $\left|D\rangle\approx|10\right\rangle$, which corresponds to the qubit's excited state.

In this parameter configuration, the qubit thus remains in the excited state even though it may be strongly coupled to the environment.
However, the readout signal, whose frequency significantly differs from the qubit and JQF frequencies, will pass the JQF and readout will not be affected.
As for the control signal applied through the waveguide, due to its high power and large photon number, the JQF becomes saturated and does not influence the qubit control.
Therefore, we can achieve high-speed readout and control of the qubit while simultaneously preserving its long lifetime.

Considering the SLH theory for cascaded quantum systems, in the reflective configuration here, the total SLH triplet is given by
\begin{equation}
G_{\rm tot}=G_{\rm j}^{\rm{L}} \triangleleft G_{\rm \phi} \triangleleft G_{\rm q}^{\rm L} \triangleleft G_{\rm 0} \triangleleft G_{\rm q}^{\rm R} \triangleleft G_{\rm \phi} \triangleleft G_{\rm j}^{\rm R},
\end{equation}
where all symbols are defined similar to those in the SLH calculation Section. 
When $\omega_{\rm j} = \omega_{\rm q} = \omega_0$ and $\phi = \omega_0/v_p\cdot\Delta x =\pi$, where $v_p$ is the phase velocity of the microwave in the waveguide, the resulting SLH triplet is:
\begin{equation}
\left\{\begin{array}{l}
\mathbf{S}=e^{i \cdot 2 \phi} = 1 \\
\mathbf{L}=\sqrt{2}\left(-\sqrt{\kappa_{\rm q}} \sigma_{\rm q}^{-}+\sqrt{\kappa_{\rm j}} \sigma_{\rm j}^{-}\right). \\
H=\omega_{\rm q} \sigma_{\rm q}^z/2+\omega_{\rm j} \sigma_{\rm j}^z/2
\end{array}\right.
\end{equation}
It is straightforward to find that there is one non-trivial dark state $|D\rangle$ for the dissipation ($\mathbf{L}|D\rangle = 0$) in the $|\rm{qubit},\rm{JQF}\rangle$ basis:
\begin{equation}
\left|D\rangle=\sqrt{\kappa_{\rm j}}|10\right\rangle+\sqrt{\kappa_{\rm q}}|01\rangle.
\end{equation}

The low-temperature measurement results for the reflective JQF Chip are presented in \figref{fig:Low temperature}.
In \figpanel{fig:Low temperature}{a}, we show the variation in the qubit relaxation lifetime $T_1$ as a function of JQF bias voltage. 
In the reflective JQF design, resonance with the qubit is achieved when this bias is set to \SI{0.231033}{\volt}. 
We compared the measured $T_1$ and $T_\phi$ of the qubit under two bias conditions, \SI{0.231033}{\volt} (on resonance) and \SI{0.20929}{\volt} (off resonance). 
Our results show that the reflective JQF design provides more than a 5-fold improvement in the qubit $T_1$. 

The final improved $T_1$ values for the reflective and giant-atom JQF configurations are similar. 
The observed qubit lifetime is slightly lower than the lifetime of over \SI{200}{\micro\second} that has been reported elsewhere for such materials~\cite{place2021new}. 
This discrepancy may be caused by the slight entanglement of the qubit with the JQF, which is smaller in size and has a larger energy participation in the lossy metal-air interface.
Additionally, we observe an enhancement in the Ramsey $T_2$ at the working point. 
Additionally, the $T_\phi$ exhibits an almost consistent value, indicating a similar amount of residual photons with frequencies close to the qubit when the JQF is tuned to or off the qubit frequency.


\subsection*{Device parameters}

\begin{table*}[htbp]
\centering
\caption{\textbf{Summary of device parameters.}}
\begin{tabular}{lccccc}
\toprule
Parameter & Symble & Qubit$_{\rm g}$ & JQF$_{\rm g}$ & Qubit$_{\rm r}$ & JQF$_{\rm r}$ \\
\midrule
Frequency (top for JQF)/\si{\mega\hertz} & $\omega_{\rm q(j)}/2\pi$ & 5011.7 & 5802 &4431.9 & 5098\\
Anharmonicity/\si{\mega\hertz} & $\alpha/2\pi$ & -302 & -348 & -270 & -344\\
Resonator frequency/\si{\mega\hertz} & $\omega_{\rm r}/2\pi$ & 6621 & - & 6497 & -\\
Qubit-resonator coupling (simulation)/\si{\mega\hertz} & $g_{\rm qr}/2\pi$ & 144 & - & 137& -\\
Resonator damping/\si{\mega\hertz} & $\kappa_{\rm r}/2\pi$ & 2.1 & -& 1.8& -  \\
JQF damping (one point)/\si{\mega\hertz} & $\kappa_{\rm j}/2\pi$ & - & 12.7  & - & 24\\
Waveguide Length & $\Delta x/(\lambda_{\rm 5GHz}/2)$ & - & 0.98 & - & 0.96\\
\bottomrule
\end{tabular}
\label{tab:qubit_parameters}
\end{table*}

The parameters of our two devices are summarized in Table~\ref{tab:qubit_parameters}. 
The terms Qubit$_g$ and JQF$_g$ refer to the giant-atom chip, the performance of which is illustrated in Fig.~4 in the main text, while Qubit$_{\rm r}$ and JQF$_{\rm r}$ denote the chip with the reflective JQF.
We normalize the waveguide length $\Delta x$ with the half-wavelength $\lambda_{\rm 5GHz}/2$ of a GHz\SI{5}{\giga\hertz} photon propagating in the waveguide. 
The length is measured from the center points of the resonator's interdigitated capacitor and the JQF coupling structure. 
Due to the heavy reliance of the qubit's coherence properties on the JQF configuration, these are not included in the table. 
When the JQF bias is not set at the optimal working point, the $T_\phi$ of Qubit$_g$ is approximately \SI{6}{\micro\second}. 
The qubit-resonator coupling $g_{\rm qr}$ and the JQF's anharmonicity are derived from simulation values, while the remaining parameters are experimentally measured.


\subsection*{SLH simulation of the effect of waveguide-length mismatch}

In this section, we employ the previously introduced SLH theory for cascaded quantum systems to investigate the variations in the $T_1$ limit associated with different off-target parameter mismatches. 
Although our numerical analysis, which is based on Qutip~\cite{johansson2012qutip}, focuses on the giant-atom-based JQF presented in the main text, the conclusions drawn are similarly applicable to the reflective JQF.

\begin{figure}[!htb]
\centering
\includegraphics[width=0.8\textwidth]{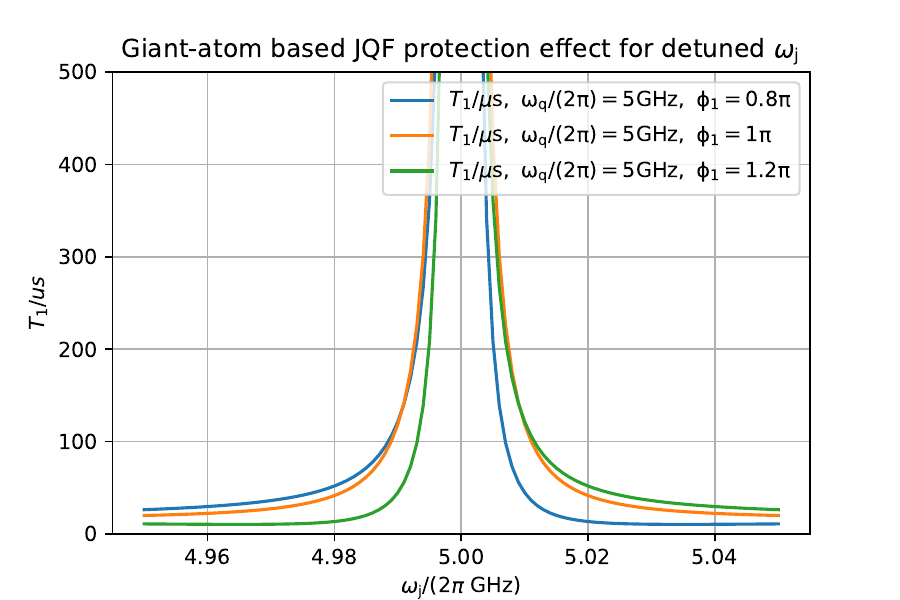}
\caption{
\textbf{Giant-atom-based JQF-limited $T_1$ for varying $\omega_{\rm j}$ and $\phi_1$.}
Qubit relaxation lifetime $T_1$ as a function of $\omega_{\rm j}$ for a set of average phases $\phi_1$, with the phase difference $\phi_2 = 0$.
}
\label{fig:wjscan}
\end{figure}

The primary parameters under investigation here are the JQF frequency $\omega_{\rm j}$, the average phase of the waveguide $\phi_1$, and the phase difference of the waveguide $\phi_2$.
In the giant-atom-based JQF, the waveguide is divided into two sections as depicted in Fig.~1 of the main text, with the phase delay of both sections defined as $\phi=\pi$ in the main text.
However, this may not accurately reflect reality, so we define the phase delay of the left-side waveguide as $\phi_l = \phi_1 + \phi_2/2$ and that of the right-side waveguide as $\phi_{\rm r} = \phi_1 - \phi_2/2$.

We first set $\phi_2=0$, vary $\phi_1=$ 0.8, 1.0, and 1.2$\times \pi$, and scan $\omega_{\rm j}$.
From the SLH triplet derived from these parameters, we can simulate the system dynamics starting from the initial qubit excited state and fit the decay of the excited-state population to obtain $T_1$ values.
The results are displayed in \figref{fig:wjscan}.
We observe an asymmetry in the $T_1$ limit profiles for different $\phi_1$ values, a phenomenon also noted in our experimental observations in Fig. 3 of the main text and Ref.~\cite{kono2020breaking}. 
The limit when $\omega_{\rm j}/2\pi = \omega_{\rm q}/2\pi =$\SI{5}{\giga\hertz} consistently remains high.

\begin{figure}[!htb]
\centering
\includegraphics[width=0.7\textwidth]{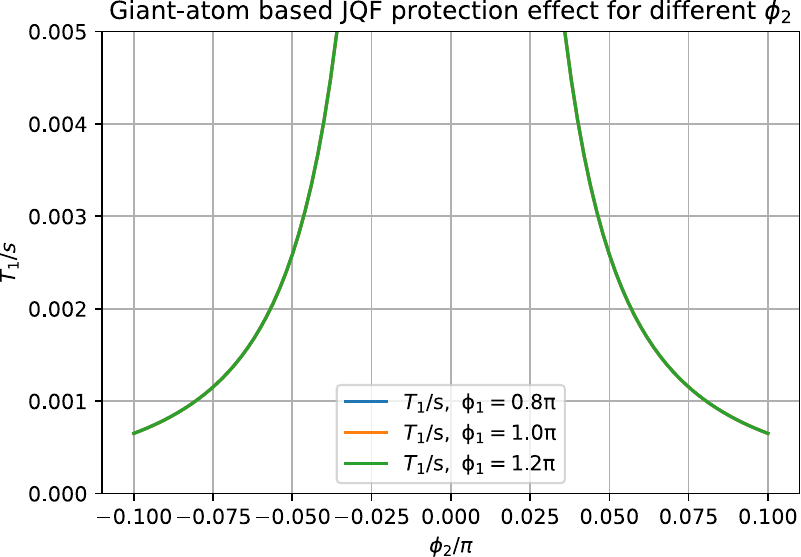}
\caption{
\textbf{Giant-atom-based JQF-limited $T_1$ for varying $\phi_2$ and different $\phi_1$.}
Qubit relaxation lifetime $T_1$ as a function of phase difference $\phi_2$. 
The results overlap for different values of $\phi_1$.  
}
\label{fig:phi2scan}
\end{figure}

Next, we fix $\omega_{\rm j}/2\pi = \omega_{\rm q}/2\pi =$\SI{5}{\giga\hertz} and investigate the effect of $\phi_2$ on $T_1$.
As we scan $\phi_2$ around 0 and simultaneously change $\phi_1$, we find that the results for different $\phi_1$ values are the same; see \figref{fig:phi2scan}.
In conclusion, the $T_1$ limit is sufficiently high ($>$0.5 ms) for $\pm 0.1\pi$ or about $\pm$ 0.9 mm in waveguide length.
This broad range of waveguide lengths facilitates the coupling of multiple qubit-resonator and JQF units, and theoretically, allows for the scaling of the giant-atom JQF structure to accommodate a multi-qubit setup.

\begin{figure}[!htb]
\centering
\includegraphics[width=0.7\textwidth]{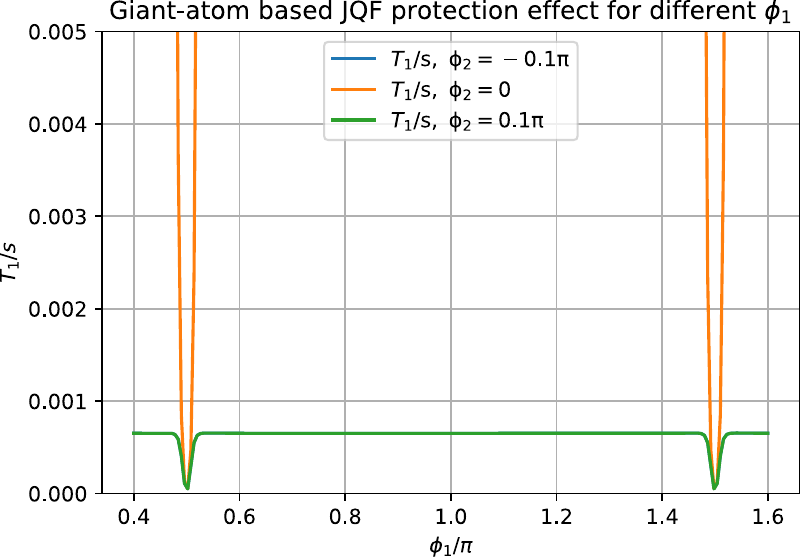}
\caption{
\textbf{Giant-atom-based JQF-limited $T_1$ for varying $\phi_1$ and different $\phi_2$.}
Qubit relaxation lifetime $T_1$ as a function of average phase $\phi_1$. 
The limitation is highest for $\phi_2=0$, while for $\phi_2 = \pm 0.1\pi$, the results overlap and the $T_1$ limit decreases, but remains at around \SI{0.7}{\milli\second}
.  
}
\label{fig:phi1scan}
\end{figure}

Now, we fix $\omega_{\rm j}/2\pi = \omega_{\rm q}/2\pi =$\SI{5}{\giga\hertz} and investigate the effect of $\phi_1$ on $T_1$.
As we scan $\phi_1$ around $\pi$ and vary $\phi_2$, we find that the results for $\phi_2=0$ are optimal; see \figref{fig:phi1scan}.
For $\phi_2 = \pm 0.1\pi$, the $T_1$ limit decreases, but it remains at around \SI{0.7}{\milli\second}, which is sufficiently high.
When the waveguide is symmetric, the protective effect of the JQF proves to be quite robust against errors in $\phi_1$.

\begin{figure}[!htb]
\centering
\includegraphics[width=\textwidth]{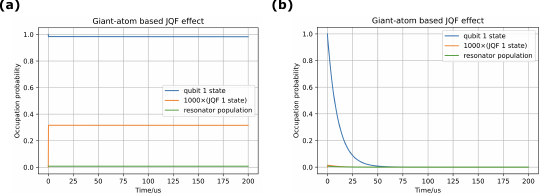}
\caption{
\textbf{Dynamics of the combined JQF-qubit-resonator system starting from the qubit excited state.}
(a) $\omega_{\rm j}/2\pi =$\SI{4.9875}{\giga\hertz}.
Given that the frequency of the qubit mode will be lower than the bare mode \SI{5}{\giga\hertz} after coupling to a \SI{6.621}{\giga\hertz} readout resonator, the operational condition for the JQF frequency $\omega_{\rm j}/2\pi=$\SI{4.9875}{\giga\hertz} is also lower than \SI{5}{\giga\hertz}.
To clearly observe the dynamics of the JQF, we have magnified its population by 1000 times in the plot.
(b) $\omega_{\rm j}/2\pi =$\SI{5.1}{\giga\hertz}.
When the JQF frequency $\omega_{\rm j}/2\pi=$\SI{5.1}{\giga\hertz} is off-resonant with the qubit mode, the JQF becomes ineffective.
The decay observed here is limited by the Purcell effect.
}
\label{fig:SLHdynamics}
\end{figure}

Finally, we incorporate the readout resonator into the SLH calculation and investigate the dynamics of this combined CQED-WQED system.
We model the JQF as a two-level system, the qubit as an anharmonic oscillator with its four lowest levels, and the resonator with its three lowest levels.
The SLH triplet now is:
\begin{equation}
\left\{\begin{array}{l}
\mathbf{S}=\left(
\begin{array}{cc}
 e^{2 i \phi _1} & 0 \\
 0 & e^{2 i \phi _1} \\
\end{array}
\right) \\
\mathbf{L}=\left(\frac{e^{i (\phi _1-\phi _2/2)} \sqrt{\kappa _{\rm r}}a+\left(1+e^{2 i \phi _1}\right) \sqrt{\kappa_{\rm j}} \sigma _{\rm j}^{-}}{\sqrt{2}},\frac{ e^{i (\phi _1+ \phi _2/2)} \sqrt{\kappa _{\rm r}}a+\left(1+e^{2 i \phi _1}\right) \sqrt{\kappa _{\rm j}} \sigma _{\rm j}^{-}}{\sqrt{2}}\right)\\
H = \begin{aligned}[t]
&\left(\omega_{\rm j}+\sin(2\phi_1)\kappa_{\rm j}\right) \sigma_{\rm j}^z/2\\
&+\omega_{\rm q}b^{\dagger}b+\alpha b^{\dagger}b^{\dagger}bb+\omega_{\rm r}a^{\dagger}a+g_{\rm qr}(ab^{\dagger}+ba^{\dagger})\\
&+\sin (\phi_1) \cos \left(\phi_2/2\right)\sqrt{\kappa_{\rm j}\kappa_{\rm r}}\left(\sigma_{\rm j}^{-} a^{\dagger}+a\sigma_{\rm j}^{+}\right).
\end{aligned}
\end{array}\right.
\end{equation}

We use $\omega_{\rm q}/2\pi =$\SI{5}{\giga\hertz}, $\phi_1 = \pi$, and $\phi_2 = 0$ for this simulation; all other parameters are consistent with those listed in Table~\ref{tab:qubit_parameters}.
The simulation results are displayed in \figref{fig:SLHdynamics}.
We observe that in order to protect the qubit from decay, the JQF also becomes populated, and initially, the qubit and JQF rapidly decay to a nontrivial dark state.
However, our qubit performance tests indicate that the minor qubit-JQF entanglement does not degrade gate and readout operations compared to an unprotected qubit. 
We also find that the JQF population changes slightly when $\phi_1$ varies, but the change is not substantial and the JQF population remains much smaller than that of the qubit.
We believe that the degree of freedom of the JQF population allows for the ultra-robust $T_1$ limit of the JQF against different $\phi_1$ values.

\end{document}